\documentclass[twocolumn,prl,superscriptaddress]{revtex4}
\usepackage{graphicx}
\usepackage{amssymb,amsmath}
\usepackage{bm}
\usepackage{dcolumn}
\usepackage{subfigure}
\usepackage{float}
\usepackage[OT1]{fontenc} 
\usepackage{url}
\usepackage{color}
\usepackage{txfonts}
\usepackage[driverfallback=dvipdfm]{hyperref}
\hypersetup{pdfpagemode=FullScreen,colorlinks=true,breaklinks,urlcolor=blue,linkcolor=blue,citecolor=blue}
\usepackage{natbib}

\usepackage{subfigure}
\usepackage{comment}
\usepackage{hyperref}
\usepackage{amssymb}
\usepackage{bm}
\usepackage{graphicx}
\usepackage{amsmath,amssymb}
\usepackage{tikz,fp}
\usepackage{tikz-cd}
\usetikzlibrary{arrows}
\usetikzlibrary{intersections}
\usetikzlibrary{shapes.geometric}
\usetikzlibrary{decorations.pathmorphing, patterns,shapes,fixedpointarithmetic}
\usetikzlibrary{decorations.markings}

\pgfdeclarepatternformonly{south west lines}{\pgfqpoint{-0pt}{-0pt}}{\pgfqpoint{3pt}{3pt}}{\pgfqpoint{3pt}{3pt}}{
        \pgfsetlinewidth{0.4pt}
        \pgfpathmoveto{\pgfqpoint{0pt}{0pt}}
        \pgfpathlineto{\pgfqpoint{3pt}{3pt}}
        \pgfpathmoveto{\pgfqpoint{2.8pt}{-.2pt}}
        \pgfpathlineto{\pgfqpoint{3.2pt}{.2pt}}
        \pgfpathmoveto{\pgfqpoint{-.2pt}{2.8pt}}
        \pgfpathlineto{\pgfqpoint{.2pt}{3.2pt}}
        \pgfusepath{stroke}}

\tikzset{
  mid arrow/.style={postaction={decorate,decoration={
        markings,
        mark=at position .585 with {\arrow{stealth}}
      }}},
  near arrow/.style={postaction={decorate,decoration={
        markings,
        mark=at position .275 with {\arrow{stealth}}
      }}},
  far arrow/.style={postaction={decorate,decoration={
        markings,
        mark=at position .800 with {\arrow{stealth}}
      }}},
  snake/.style={fixed point arithmetic, decorate, decoration={snake,amplitude=2pt, segment length=6pt}},
}

\begin{document}
  
  \title{Universal subdiffusion in strongly tilted many-body systems }
  \author{Pengfei Zhang}
  \affiliation{Institute for Quantum Information and Matter and Walter Burke Institute for Theoretical Physics, California Institute of Technology, Pasadena, California 91125, USA}
  \date{\today}
  
  \begin{abstract}
  The quantum dynamics away from equilibrium is of fundamental interest for interacting many-body systems. In this letter, we study tilted many-body systems using the effective Hamiltonian derived from the microscopic description. We first give general arguments for the density relaxation rate satisfying $1/\tau\propto k^4$ for a large class of systems, including the Fermi Hubbard model case as observed in the the recent experiment \cite{guardado2020subdiffusion}. Here $k$ is the wave vector of the density wave. The main ingredients are the emergence of the reflection symmetry and dipole moment conservation to the leading non-trivial order of the large tilted strength. To support our analysis, we then construct a solvable model with large local Hilbert space dimension by coupling sites discribed by the Sachdev-Ye-Kitaev models, where the density response can be computed explicitly. The the tilt strength and the temperature dependence of the subdiffusion constant are also discussed. 
  \end{abstract}
  
  \maketitle
  \textit{Introduction} In recent years, much understanding of strongly interacting systems has been gained thanks to the quantum simulation with cold atom systems. By using the quantum gas microscope for fermions \cite{haller2015single,cheuk2015quantum,edge2015imaging,omran2015microscopic,greif2016site,cheuk2016observation2}, the antiferromagnetic correlation \cite{parsons2016site,boll2016spin,cheuk2016observation,hilker2017revealing} has been observed in the Fermi Hubbard model, which is suspected to be related to the understanding of the high-temperature superconductivity. Besides observables in thermal ensembles, non-equilibrium quantum dynamics can also be used as a probe of underlying physics. As an example, the spin \cite{nichols2019spin} and charge \cite{brown2019bad} diffusion have been studied experimentally, which shows the bad-metal behavior. Except for these standard transport experiments, the precise control of the Fermi Hubbard model in optical lattice also paves a new way to study physics in some extreme conditions. In a recent work \cite{guardado2020subdiffusion}, by applying an additional laser beam, authors realize a strong linear potential, or tilt, in the 2D Fermi Hubbard model. This is an analogy of applying a strong electric field in solid-state systems. A crossover between a traditional diffusion in the weak tilt limit \cite{mandt2011interacting} and a new subdiffusion behavior in the strong tilt limit has been observed. 

  In this work, we find this subdiffusion behavior is universal for a large class of tilted systems. Using the large tilt strength expansion, we obtain the low-energy effective Hamiltonian which governs the dynamics of the strongly tilted system. The subdiffusion behavior is then found to be a direct consequence of the emergent reflection symmetry and the dipole moment conservation. We also analyze the scaling of the subdiffusion constant with respect to the tilt strength. The analysis is under the assumption of hydrodynamical behavior, which is motivated by the experiments. Note that this is different from the low-dimensional models \cite{sala2020ergodicity,van2019bloch,schulz2019stark,morningstar2020kinetically} which show transitions between ergodic and nonergodic phases. Very recently, subdiffusion behavior is also observed in a specific circuit model that conserves the dipole momentum \cite{feldmeier2020anomalous}, where both symmetries are imposed from the beginning \footnote{In this paper, authors also study the consequence of the higher-moment conservation.}, and in more sophisticated fracton models \cite{pretko2020fracton,gromov2020fracton}. To further support our analysis, we construct a solvable model by adding strong tilted potential to the quadratically coupled Sachdev-Ye-Kitaev model (SYK) \cite{kitaev2014hidden,maldacena2016remarks,zhang2017dispersive,song2017strongly}.  Without the tilted potential, the model is a strongly interacting non-Fermi liquid with diffusive charge transport at finite temperature \cite{song2017strongly}, which provides an ideal platform for our study. We explicitly compute the density response in the leading order of large tilted potential which shows the expected subdiffusion behavior. 
  \begin{figure}[t]
  \centering
  \includegraphics[width=1\columnwidth]{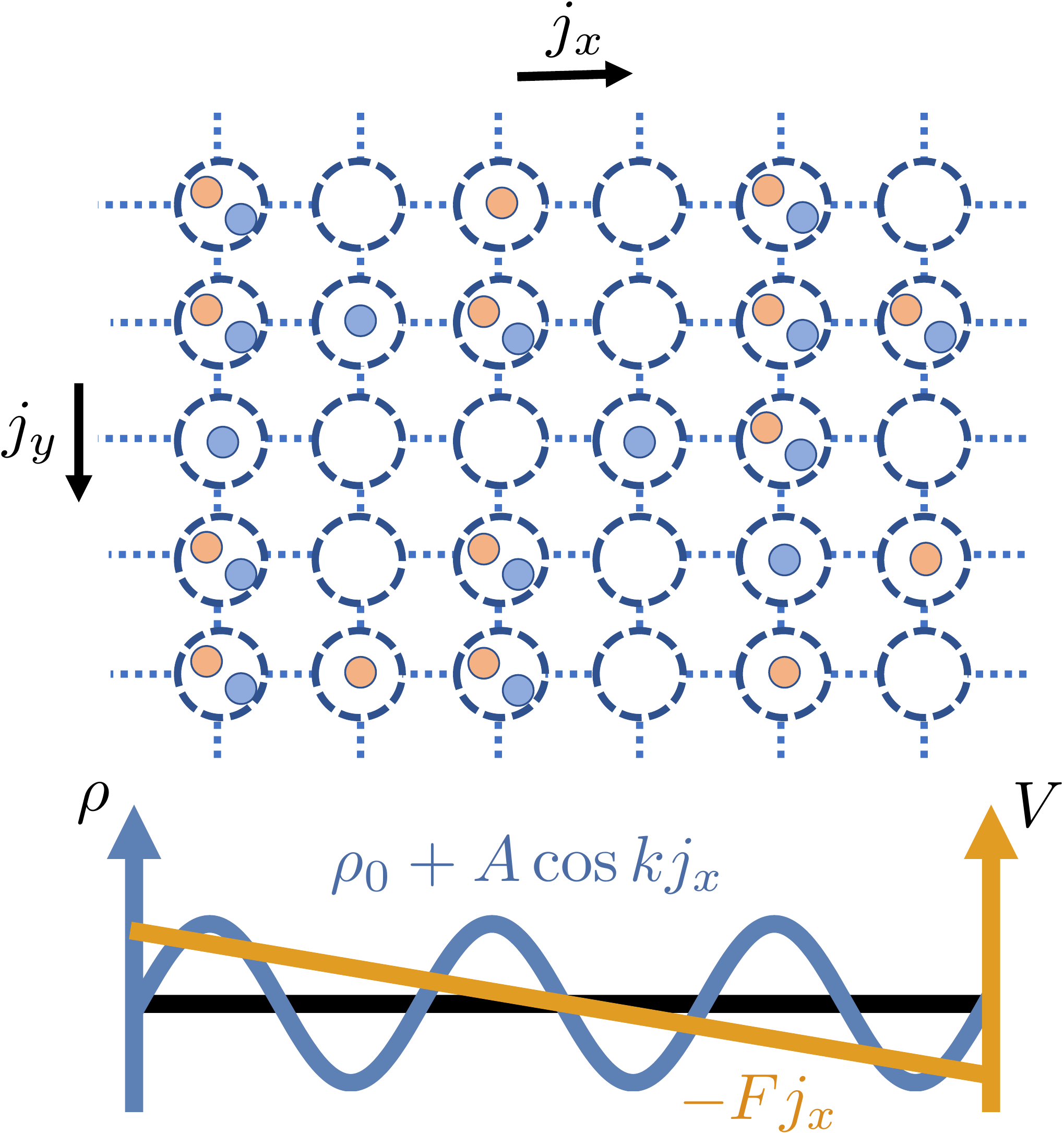}
  \caption{A schematic of the tilted models on a two-dimensional square lattice. $J$ denotes the hopping strength and $F$ is the slope of the linear potential. Balls with different colors indicate atoms with different internal states. We focus on the relaxation of the initial density wave $\rho(j_x,t)=\rho_0+A(t)\cos(k j_x)$. }\label{fig:schematic}
\end{figure}

  \textit{General Analysis} In this work, we consider quantum many-body systems with additional tilted potential. To be concrete, we focus on the two-dimensional square lattice with nearest-neighbor hopping and on-site interactions, as illustrated in Fig. \ref{fig:schematic}. The Hamiltonian reads 
  \begin{equation}\label{H}
  \begin{aligned}
  H=&-\sum_{\mathbf{j},a}\left(J_xc^\dagger_{\mathbf{j}+\hat{x},a}c_{\mathbf{j},a}+J_yc^\dagger_{\mathbf{j}+\hat{y},a}c_{\mathbf{j},a}+\text{H.C.}\right)\\&+\sum_\mathbf{j} H^\text{int}(c^\dagger_{\mathbf{j},a},c_{\mathbf{j},a})-\sum_{\mathbf{j}} F j_xn_{\mathbf{j}}.
  \end{aligned}
  \end{equation}
  Here $c_{\mathbf{j},a}$ ($c_{\mathbf{j},a}^\dagger$) is the annihilation (creation) operator on site $\mathbf{j}=(j_x,j_y)$ with additional internal-state index $a$. We have set the lattice constant $a_{\text{Lattice}}=1$ and assumed the interactions preserve the local particle number $n_{\mathbf{j}}\equiv\sum_{a}n_{\mathbf{j},a}=\sum_{a}c_{\mathbf{j},a}^\dagger c_{\mathbf{j},a}$. For the Fermi-Hubbard case \cite{guardado2020subdiffusion}, we have $a\in\{\uparrow,\downarrow\}$ and $H^\text{int}(c^\dagger_{\mathbf{j},a},c_{\mathbf{j},a})=Un_{\mathbf{j},\uparrow}n_{\mathbf{j},\downarrow}$. The last term in \eqref{H} presents a tilted potential in the $x$ direction. Note that although in this letter we only give explicit examples in term of fermions, the general argument presented below also applies for interacting bosons or boson-fermion mixtures.

  We are interested in charge diffusion dynamics $\rho(\mathbf{j},t)=\left<\psi|n_{\mathbf{j}}(t)|\psi\right>$ with some initial state $ |\psi \rangle$. For initial configurations with density waves $\rho(j_x,t)=\rho_0+A(t)\cos(k j_x)$, we expect its amplitude decays exponentially as $A(t)\propto e^{-t/\tau}$ for general interacting systems \footnote{Note that this is an important assumption, which excludes the possibility of non-ergodicity. }. We would like to understand the relation between $\tau$ and the wave vector $k$ of the density wave. To analyze this problem, we first consider applying a gauge transformation $\mathcal{U}=\exp(i\sum_{\mathbf{j}} F j_xn_{\mathbf{j}} t)$. The Hamiltonian becomes
  \begin{equation}
  \begin{aligned}
  \tilde{H}(t)=&-\sum_{\mathbf{j},a} \left(J_xe^{-iF t}c^\dagger_{\mathbf{j}+\hat{x},a}c_{\mathbf{j},a}+J_yc^\dagger_{\mathbf{j}+\hat{y},a}c_{\mathbf{j},a}+\text{H.C.}\right)\\&+\sum_\mathbf{j} H^\text{int}(c^\dagger_{\mathbf{j},a},c_{\mathbf{j},a}).
  \end{aligned}
  \end{equation}
  $\tilde{H}(t)$ is time-periodic with frequency $F$. For $F$ much larger than any other energy scales, one can make a high-frequency expansion to obtain a Floquet effective Hamiltonian, which governs the dynamics of the system at integer times of the driven period $T=2\pi/F$. The Floquet effective Hamiltonian $H_{\text{eff}}$ is defined as
  \begin{equation}
 U(T)=\mathcal{T}\exp\left(-i\int^{T}_0\tilde{H}(t)dt\right)=  \exp\left(-iH_{\text{eff}}T\right).
  \end{equation}
Here $\mathcal{T}$ is the real-time ordering operator. Defining $\tilde{H}(t)=\sum_{m=0,\pm1} H_m e^{imF t}$ with
\begin{equation}
\begin{aligned}
H_0=&-J_y\sum_{\mathbf{j},a}\left(c^\dagger_{\mathbf{j}+\hat{y},a}c_{\mathbf{j},a}+\text{H.C.}\right)+\sum_\mathbf{j} H^\text{int}(c^\dagger_{\mathbf{j},a},c_{\mathbf{j},a}),\\
H_{1}=&-J_x\sum_{\mathbf{j},a}c^\dagger_{\mathbf{j},a}c_{\mathbf{j}+\hat{x},a}=H_{-1}^\dagger,
\end{aligned}
\end{equation}
the effective Hamiltonian reads \cite{eckardt2017colloquium}:
\begin{equation}\label{Heff}
H_{\text{eff}}\equiv H_0+\Delta H=H_0+ \frac{[H_{-1},[H_0,H_1]]}{F^2}.
\end{equation}
Here we keep terms to the order of $(|H|/F)^2$, with $|H|$ being the typical energy scale of hopping or interaction terms. Importantly, the term proportional to $1/F$ vanishes since $[H_1,H_{-1}]=0$ and the leading order non-trivial correlation between sites with different $x$ comes from the $1/F^2$ term. It is also straightforward to show that only the interaction term in $H_0$ contributes to $\Delta H$. However, for the present discussion, the specific form of this term is not significant.

For strongly interacting systems, it is plausible to assume the evolution of (coarse-grained) density $\rho(\mathbf{x},t)$ can be described by the hydrodynamical equation $\partial_t\rho(\mathbf{x},t)+\nabla \cdot \mathbf{J}(\mathbf{x},t)=0$ in the long-time limit. The existence of the subdiffusion behavior (in $x$ direction) in general interacting many-body systems with large tilted potential comes from two observations:

Firstly, the system has the translation symmetry and an emergent reflection symmetry $\mathcal{P}$, which generates $x\rightarrow -x$. This can be understood by realizing that the microscopic Hamiltonian \eqref{H} satisfies 
\begin{align}
\mathcal{P}H(F)\mathcal{P}=H(-F).
\end{align}
Consequently, we have an emergent reflection symmetry $\mathcal{P}H_{\text{eff}}\mathcal{P}=H_{\text{eff}}$ since the effective Hamiltonian only depends on $F^2$. This implies that if take a gradient expansion of $J_x$ in terms of $\rho$, it should only contain an odd number of derivatives
\begin{equation}\label{expansion}
J_x(\mathbf{x},t)=-D^{(1)}\partial_x\rho(\mathbf{x},t)+D^{(3)}\partial_x^3\rho(\mathbf{x},t)+...
\end{equation} 
Here $D^{(1)}$ is the usual charge diffusion constant, while $D^{(3)}$ is the subdiffusion constant.

Secondly, the effective Hamiltonian conserves the dipole moment $\mathcal{D}=\sum_{\mathbf{j}}j_xn_{\mathbf{j}}$:
\begin{equation}
[\mathcal{D},H_{\text{eff}}]=0,
\end{equation}
since $[\mathcal{D},H_0]=0$ and $[\mathcal{D},H_{\pm1}]=\mp H_{\pm1}$. This emergence of dipole moment conservation can also be directly understood as taking the large $F$ limit of \eqref{H}. As a result, we expect local conservation of dipole moment
\begin{equation}
\partial_t d=x\partial_xJ_x=\partial_x \left[xJ_x-D^{(1)}\rho+D^{(3)}\partial_x^2\rho\right]\equiv\partial_x J^d_x.
\end{equation}
Here we consider only variations in the $x$ direction. $d(x,t)=x \rho(x,t)$ is the local dipole moment with current $$J^d_x=xJ_x-D^{(1)}\rho+D^{(3)}\partial_x^2\rho.$$ Since the Hamiltonian is translation invariant, we expect $J^d_x=0$ for a homogeneous charge filling. As a result, we should have $D^{(1)}=0$ and the hydrodynamical equation writes
\begin{equation}\label{hy}
\partial_t\rho(x,t)+D^{(3)}\partial_x^4\rho=0,
\end{equation}
which leads to the relaxation rate $1/\tau \propto k^4$. Similar subdiffusion equation has also been found recently in fracton models \cite{gromov2020fracton} and specific circuit model with dipole moment conservation \cite{feldmeier2020anomalous}. Our analysis here can be directly applied to the experimental case of the Fermi Hubbard model \cite{guardado2020subdiffusion}. Since the analysis is based on the symmetry of the system and does not depend on the details of the local interaction term, the existence of subdiffusion behavior should be general: It is straightforward to generalize the result into a higher dimension, different lattices with reflection symmetry and different interaction terms that commute with the gauge transformation $\mathcal{U}$.  We could also consider microscopic Hamiltonians that explicitly break the reflection symmetry when $F=0$. In this case, it is possible that the charge relaxation behavior is modified since generally additional terms $\sim \partial_x^2 \rho$ may appear in \eqref{expansion}. 

Moreover, we could determine the scaling of $D^{(3)}$ with respect to the tilt strength $1/F$. The correlation in $x$ direction is from $\Delta H \sim UJ_x^2/F^2$, where $U$ is the typical interaction strength \footnote{We have used the fact that the hopping term in $H_0$ commutes with $H_1$.}. If we consider a transition driven by this term between two different charge configurations, the transition rate $R$ is then proportional to $ 1/F^4$. For small interaction strength, one further expects $R\propto U^2/F^4$. It is reasonable to assume $D^{(3)}$ show the same scaling as $R$. 

\textit{Solvable Model} 
Although our argument applies for general interacting quantum systems, a determination of the subdiffusion constant is usually hard and involves additional approximations. Here, to support our argument, we instead consider specific local interaction terms with which the model can be solved explicitly with non-trivial interactions. 

The SYK model \cite{kitaev2014hidden,maldacena2016remarks}, which describes $N$ random interacting fermion modes in $0+1$-D, is known as a solvable model in the large-$N$ limit where the dynamical mean-field theory \cite{georges1996dynamical} becomes exact. The model describes a strongly correlated non-Fermi liquid with maximal many-body chaos. Generalizations with spatial dimension can then be systematically constructed by coupling different SYK sites \cite{banerjee2017solvable,chen2017competition,zhang2017dispersive,jian2017solvable,gu2017local,davison2017thermoelectric}. For generalizations with $U(1)$ symmetry \cite{davison2017thermoelectric,gu2020notes}, the charge transport shows the standard diffusive behavior. As a result, it provides an ideal platform to study the effect of a large tilted potential. 

To construct a coupled SYK system with additional tilt, we consider a fermionic system with large number of internal states $a\in \{1,2...N\}$. We further specify the on-site interaction $H^\text{int}$ in \eqref{H} as 
\begin{equation}
H^\text{int}(c^\dagger_{\mathbf{j},a},c_{\mathbf{j},a})=\frac{1}{4}\sum_{abcd}U^\mathbf{j}_{ab;cd}c^\dagger_{\mathbf{j},a}c^\dagger_{\mathbf{j},b}c_{\mathbf{j},c}c_{\mathbf{j},d}.
\end{equation}
Here $U^\mathbf{j}_{ab;cd}$ with different indices are independent random Gaussian variables with expectation and variance
\begin{equation}
\overline{U^\mathbf{j}_{ab;cd}}=0,\ \ \ \ \ \ \overline{\left(U^\mathbf{j}_{ab;cd}\right)^2}=\frac{2U^2}{N^3},
\end{equation}
under the symmetry $U^\mathbf{j}_{ab;cd}=-U^\mathbf{j}_{ba;cd}=-U^\mathbf{j}_{ab;dc}$. The system is translational invariant after the disorder average. We assume the system is self-averaged and take the disorder replica diagonal assumption \cite{maldacena2016remarks}. For $F=J_x=J_y=0$, the Hamiltonian \eqref{H} describes decoupled SYK models. Near this SYK fixed point, the hopping term is relevant which leads to a crossover to Fermi liquid behavior at $T \sim \text{max}\{J_x^2/U,J_y^2/U\}$ \cite{chen2017competition,song2017strongly}.

We now consider the effect of a strong tilted potential by evaluating the effective action \eqref{Heff} explicitly. We find
\begin{equation}
\begin{aligned}
\Delta H=&\frac{J_x^2}{4F^2}\sum_{abcd,\mathbf{j}}U^\mathbf{j}_{ab;cd}C^\mathbf{j}_{ab;cd},\\
C^\mathbf{j}_{ab;cd}=&-4c^\dagger_{\mathbf{j},a}c^\dagger_{\mathbf{j},b}c_{\mathbf{j},c}c_{\mathbf{j},d}-2c^\dagger_{\mathbf{j}+\hat{x},a}c^\dagger_{\mathbf{j}-\hat{x},b}c_{\mathbf{j},c}c_{\mathbf{j},d}\\
&-2c^\dagger_{\mathbf{j},a}c^\dagger_{\mathbf{j},b}c_{\mathbf{j}-\hat{x},c}c_{\mathbf{j}+\hat{x},d}+4c^\dagger_{\mathbf{j}+\hat{x},a}c^\dagger_{\mathbf{j},b}c_{\mathbf{j},c}c_{\mathbf{j}+\hat{x},d}\\
&+4c^\dagger_{\mathbf{j}-\hat{x},a}c^\dagger_{\mathbf{j},b}c_{\mathbf{j},c}c_{\mathbf{j}-\hat{x},d}
\end{aligned}
\end{equation}
The first term gives a renormalization of the local interaction $U^\mathbf{j}_{ab;cd}$. Other terms describe interactions between nearest sites and collective hopping that conserves the dipole momentum. 

We first consider the two-point function $G_{\mathbf{j}}(\tau)=\left<\mathcal{T}_\tau c_{\mathbf{i},a}(\tau)c^\dagger_{\mathbf{0},a}(0)\right>$ where $\mathcal{T}_\tau$ is the imaginary-time ordering operator. Using melon diagrams, it is straightforward to show that a self-consistent solution can be derived by assuming the Green's function vanishes unless $j_x=0$, similar to the case in \cite{gu2017local}. This leads to the self-consistent equation:
\begin{equation}
\begin{aligned}
&\Sigma(\tau)=\begin{tikzpicture}[baseline={([yshift=-6pt]current bounding box.center)}, scale=1.1]
\draw[thick,mid arrow] (-24pt,0pt) -- (-13pt,0pt);
\draw[dashed,thick] (-13pt,0pt)..controls (-8pt,18pt) and (8pt,18pt)..(13pt,0pt);
\draw[thick,mid arrow] (-13pt,0pt)..controls (-8pt,-10pt) and (8pt,-10pt)..(13pt,0pt);
\draw[thick,mid arrow] (-13pt,0pt)..controls (-8pt,10pt) and (8pt,10pt)..(13pt,0pt);
\draw[thick,mid arrow] (13pt,0pt) -- (-13pt,0pt);
\draw[thick,mid arrow] (13pt,0pt) -- (24pt,0pt);
\end{tikzpicture}
=U^2\left(1-\frac{8J_x^2}{F^2}\right)G_\mathbf{0}^3(\tau),\\
&G_{\mathbf{k}}(\omega)=(-i\omega_n-2J_y \cos(k_y)-\Sigma(\omega))^{-1}.
\end{aligned}
\end{equation}
Here the dashed line represents the disorder average. We have introduced the $G_{\mathbf{j}}(\tau)=\frac{1}{\beta}\sum_n\int \frac{d\mathbf{k}}{2\pi}G_{\mathbf{k}}(\omega)e^{i\mathbf{k}\cdot \mathbf{j}-i\omega_n t}$ with Matsubara frequency $\omega_n=(n+1/2)\pi/\beta$. We have kept terms up to the $J_x^2/F^2$ order. For $J_y/UT\ll1$ and $\beta U \gg 1$, the system has emergent conformal symmetry where the scaling dimension of $c$ is $1/4$, as in the original SYK model. This leads to the Green's function of fermions 
\begin{equation}\label{G}
G_{\mathbf{j}}(\tau)=\frac{\text{sgn}(\tau)}{(4\pi \tilde{U}^2)^{1/4}}\left[\frac{\pi}{\beta\sin (\pi\tau/\beta) }\right]^{1/2}\delta_{\mathbf{j}{\mathbf{0}}}.
\end{equation}
Here $\tilde{U}=U(1-\frac{4J_x^2}{F^2})$. At $J_y/UT\sim1$, the system exhibits a crossover to a stack of Fermi liquid chains. Note that for $J_y\ll J_x$, this crossover temperature can be much smaller than the $F=0$ case.

We now study the charge fluctuation in $x$-direction. We follow the idea in \cite{song2017strongly,zhang2017dispersive} by using the Keldysh approach and introducing considering the phase fluctuation in $x$-direction. The Keldysh contour \cite{kamenev2011} contains a forward and a backward evolution, with fermion field $c_{\mathbf{j},a,\pm}(t)$. The path-integral writes
\begin{equation}
\begin{aligned}
 Z&=\int \mathcal{D}c_{\mathbf{j},a,m}\mathcal{D}c_{\mathbf{j},a,m}^\dagger\exp(iS_K),\\
S_K&=\sum_m\int_{-\infty}^\infty dt\ m\left[\sum_{\mathbf{j}a}c_{\mathbf{j},a,m}^\dagger(i\partial_t)c_{\mathbf{j},a,m}-H_{\text{eff}}(c^\dagger_{\mathbf{j},a,m},c_{\mathbf{j},a,m})\right],
\end{aligned}
\end{equation}
with additional boundary terms at $t=\pm \infty$ imposing the fluctuation-dissipation theorem. Here we have defined $m=\pm$. The real-time Green's function on the Keldysh contour can be directly related to the imaginary-time Green's function $G_{\mathbf{j}}(\tau)$ by the analytical continuations.

The phase fluctuation can be introduced as $$c_{\mathbf{j},a,\pm}(t)\rightarrow e^{-i\phi_\pm(x,t)}c_{\mathbf{j},a}(t).$$ Here $\phi_\pm(x)$ is a slow-varying phase factor. The non-trivial coupling terms between $\phi$ and fermions read 
\begin{equation}
\begin{aligned}
S^c_K=&\sum_m\int dt\ m\left[\sum_{\mathbf{j},a}c_{\mathbf{j},a,m}^\dagger(\partial_t\phi_m)c_{\mathbf{j},a,m}+\sum_{\mathbf{j},abcd}\frac{J_x^2U^\mathbf{j}_{ab;cd}}{2F^2}\right.\\
&\left. \times \left(c^\dagger_{\mathbf{j}+\hat{x},a,m}c^\dagger_{\mathbf{j}-\hat{x},b,m}c_{\mathbf{j},c,m}c_{\mathbf{j},d,m}e^{i\partial_x^2\phi_m}+\text{H.C.}\right)\right].
\end{aligned}
\end{equation}
Since we focus on the density response $\Pi_{R}(t,\mathbf{j})=-i\theta(t)[n_\mathbf{j}(t),n_\mathbf{0}]$. We further introduce source term:
\begin{equation}
S^s_K=\sum_m\int dt\ m\left[\sum_{\mathbf{j}}n_{\mathbf{j},m}\mathcal{J}_{\mathbf{j},m}\right].
\end{equation}
which leads to $\Pi_R=-\frac{i}{2}\frac{\partial^2\log Z}{\partial \mathcal{J}_{\text{cl}}\partial \mathcal{J}_{\text{q}}}$ with $O_{\text{cl}/\text{q}}\equiv (O_+ \pm O_-)/2$. 
To calculate $\Pi_{R}$ to the leading order of $1/N$, we keep the action of $\phi$ to the quadratic order \cite{song2017strongly,zhang2017dispersive}. The action takes the form: 
\begin{equation}\label{effphi}
\begin{aligned}
\frac{iS^{\phi}_{\text{q,cl}}[\phi,J]}{N}&=
\begin{tikzpicture}[baseline={([yshift=-1pt]current bounding box.center)}, scale=1.1]
\draw[thick,snake] (-19pt,0pt) -- (-10pt,0pt);
\draw[thick,mid arrow] (-10pt,0pt)..controls (-6pt,-12pt) and (6pt,-12pt)..(10pt,0pt);
\draw[thick,mid arrow] (10pt,0pt)..controls (6pt,12pt) and (-6pt,12pt)..(-10pt,0pt);
\draw[thick,snake] (10pt,0pt) -- (19pt,0pt);
\end{tikzpicture}+
\begin{tikzpicture}[baseline={([yshift=-6pt]current bounding box.center)}, scale=1.1]
\draw[thick,snake] (-22pt,0pt) -- (-13pt,0pt);
\draw[dashed,thick] (-13pt,0pt)..controls (-8pt,18pt) and (8pt,18pt)..(13pt,0pt);
\draw[thick,mid arrow] (-13pt,0pt)..controls (-8pt,-10pt) and (8pt,-10pt)..(13pt,0pt);
\draw[thick,mid arrow] (13pt,0pt)..controls (8pt,10pt) and (-8pt,10pt)..(-13pt,0pt);
\draw[thick,mid arrow] (-13pt,0pt)..controls (-8pt,-3.5pt) and (8pt,-3.5pt)..(13pt,0pt);
\draw[thick,mid arrow] (13pt,0pt)..controls (8pt,3.5pt) and (-8pt,3.5pt)..(-13pt,0pt);
\draw[thick,snake] (13pt,0pt) -- (22pt,0pt);
\end{tikzpicture}+
\begin{tikzpicture}[baseline={([yshift=-6pt]current bounding box.center)}, scale=1.1]
\draw[dashed,thick] (0pt,-10pt)..controls (-18pt,-6pt) and (-14pt,6pt)..(0pt,10pt);
\draw[thick,mid arrow] (0pt,-10pt)..controls (-10pt,-6pt) and (-10pt,6pt)..(0pt,10pt);
\draw[thick,mid arrow] (0pt,-10pt)..controls (-3.5pt,-6pt) and (-3.5pt,6pt)..(0pt,10pt);
\draw[thick,mid arrow] (0pt,10pt)..controls (10pt,6pt) and (10pt,-6pt)..(0pt,-10pt);
\draw[thick,mid arrow] (0pt,10pt)..controls (3.5pt,6pt) and (3.5pt,-6pt)..(0pt,-10pt);
\draw[thick,snake] (-12pt,-14pt) -- (0pt,-10pt);
\draw[thick,snake] (12pt,-14pt) -- (0pt,-10pt);
\end{tikzpicture}\\
&=\sum_{\omega,k_x}\Lambda_1(\omega)(\partial_t\phi_\text{cl}+\mathcal{J}_\text{cl})(\omega,k_x)(\partial_t\phi_\text{q}+\mathcal{J}_\text{q})(-\omega,-k_x)\\
&\ \ \ +k_x^4(\Lambda_2(\omega)-\Lambda_2(0))\phi_\text{cl}(\omega,k_x)\phi_\text{q}(-\omega,-k_x).
\end{aligned}
\end{equation}
Here the wavy line represents the phase fluctuation $\phi$ or source $\mathcal{J}$. Focusing on the retarded part, we only maintain terms contain both the classical and the quantum components. The first term describes the response of a homogeneous source field. As a result in the low-energy limit $\Lambda_1(\omega\rightarrow 0)\rightarrow2iK$ with $KN$ being the compressibility \cite{gu2020notes}. The other coefficient writes:
\begin{equation}
\begin{aligned}
\Lambda_2(t)=-\frac{J_x^4}{2F^4}&\left(G_R(t)\Sigma_K(-t)+G_A(-t)\Sigma_K(-t)\right.
\\&\left.+\Sigma_R(t)G_K(-t)+\Sigma_A(-t)G_K(-t)\right).
\end{aligned}
\end{equation} 
Here we have defined the retarded Green's function $G_{R/A}(t)=-i\theta(\pm t)\{c^\dagger_{\mathbf{j},a}(t),c_{\mathbf{j},a}(0)\}$ with $\Sigma_{R/A}(t)$ being corresponding self-energy. The Keldysh component reads $G_K(\omega)=(G_R(\omega)-G_A(\omega))(1-2n_F(\omega))\equiv-2\pi iA_G(\omega)(1-2n_F(\omega))$ and similarly for $\Sigma_K$. For small $\omega$, we expand $\Lambda_2(\omega)-\Lambda_2(0)\approx-2\omega K D^{(3)}$ with
\begin{equation}
KD^{(3)}=-\frac{\pi J_x^4}{F^4}\int dq_0 A_G(q_0)A_\Sigma(q_0)\partial_{q_0}n_F(q_0)\approx \frac{J_x^4}{2F^4}.
\end{equation}
We have used the conformal Green's function \eqref{G}. Using the effective action \eqref{effphi}, the density response can be computed as:
\begin{equation}
\Pi_R(\omega,k_x)=\frac{KND^{(3)}k_x^4}{-i\omega+D^{(3)}k_x^4}.
\end{equation}
The pole $-i\omega=-D^{(3)}k_x^4$ leads to the correct hydrodynamical behavior as in \eqref{hy}. We see the subdiffusion constant $D^{(3)}$ has a scaling of $1/F^4$ consistent with our previous analysis. Interestingly, the factor of $U$ in $KD^{(3)}$ cancels out, which is distinct from the expectation for weakly interacting systems.

We finally comment on the different temperature dependence of the (sub)diffusion constant compared to the result with $F=0$, as computed in \cite{song2017strongly}. Without the tilted potential, the diffusion constant is proportional to $1/T$, leading to a linear resistivity as in strange metals. This is because the correlation between different sites is from a quadratic term, which is relevant at the low energy limit with a scaling dimension $1/2$. However, here the correlation is from a term that is marginal, leading to a constant subdiffusion constant. More generally, we could consider generalizing the local interaction term to be the SYK$_q$ model. While scaling of the diffusion constant with $F=0$ changes, the subdiffusion constant for the strongly tilted potential remains a constant. In this sense, we expect a weaker temperature dependence for $D^{(3)}$ in strongly correlated systems.

\textit{Outlook} In this work we have analyzed quantum many-body systems in strongly tilted potential. We find the charge transport show subdiffusion behavior due to the emergent reflection symmetry and dipole momentum conservation. We further determine the subdiffusion constant in a solvable coupled SYK model as an explicit example. 

Several interesting questions can be asked along this direction. Firstly, we could consider generalizing the tilted potential to impose dynamical conservation of high-order moments. However, this would break the translation symmetry and make the analysis harder. Secondly, here we have considered the case where $F$ is the largest energy scale of the system. We could ask what happens if we instead consider the near-resonant case with $F\approx U$ in the Hubbard model case \cite{bukov2016schrieffer,itin2015effective,gorg2018enhancement,xu2018correlated,sun2019resonant}. It is also interesting to study the possible interplay between thee spin diffusion and the charge diffusion in the strongly tilted systems. We postpone all these questions for future studies.

\textit{Acknowledgment} We would like to thank Lei Pan for bringing our attention to the experiment \cite{guardado2020subdiffusion}. 
\bibliography{ref.bib}

\end{document}